# On the ballistic transport in nanometer-scaled double-gate MOSFET

Jérôme Saint Martin, Arnaud Bournel, Philippe Dollfus

*Institut d'Electronique Fondamentale, CNRS UMR 8622, Université Paris Sud, Bât. 220, F-91405 Orsay cedex, France*

**Corresponding author**

Jérôme Saint Martin
Institut d'Electronique Fondamentale
CNRS UMR 8622, Université Paris Sud, Bât. 220
F-91405 Orsay cedex, France
Tel.: 33 1 69 15 40 37, Fax: 33 1 69 15 40 20, E-mail: stmartin@ief.u-psud.fr

**Abstract**

The scattering effects are studied in nanometer-scaled double-gate MOSFET, using Monte Carlo simulation. The non-equilibrium transport in the channel is analyzed with the help of the spectroscopy of the number of scatterings experienced by electrons. We show that the number of ballistic electrons at the drain-end, even in terms of flux, is not the only relevant characteristic of ballistic transport. Then the drive current in the 15 nm-long channel transistor generations should be very close to the value obtained in the ballistic limit even if all electrons are not ballistic. Additionally, most back-scattering events which deteriorates the ON current, take place in the first half of the channel and in particular in the first low field region. However, the contribution of the second half of the channel can not be considered as negligible in any studied case i.e. for a channel length below 25 nm. Furthermore, the contribution of the second half of the channel tends to be more important as the channel length is reduced. So, in ultra short channel transistors, it becomes very difficult to extract a region of the channel which itself determine the drive current $I_{on}$.

# 1 Introduction

Double-gate MOSFET architecture (DGMOS) is a potential solution to overcome short channel effects in the 65 nm ITRS node [1], that is for physical gate lengths smaller than 25 nm. In such nano-transistors where the channel length is comparable to the electron mean free path, non stationary [2] or even ballistic [3] transport is probably of great importance regarding the device performance. This question is however rather controversial from a theoretical point of view.

According to Natori prediction [4], non stationary transport in ultra-small devices and statistical fluctuation of random scattering events undergone by charge carriers in the channel should lead to dramatic time fluctuations of drive current. According to Monte Carlo simulation results, small variations in the number and the position of doping atoms in the channel of a 50 nm bulk MOSFET significantly influence the transport properties and the drain current [5]. Ballistic transport in undoped channel may be the solution to limit these types of fluctuations.

Lundstrom and coworkers have studied in detail the influence of ballistic transport in such devices on the drain current [6],[7]. They developed models describing nano-DGMOS operating in the ballistic or quasi-ballistic limit. These models are based on the concept of thermionic injection from source-end into the channel. According to such approach, the electron velocity at the source-end, and thus the drive current, should be limited by the "source-to-channel" energy barrier and by back-scatterings in the low-field region, i.e. in the vicinity of the barrier.

Svizhenko and Anantram have also investigated the role of scatterings in nanometer-scaled DGMOS [8]. Using a Green-function approach, they show that scatterings at both source- and drain-end influence significantly the drain current.

More recently, Mouis and Barraud [9] have used the Monte Carlo simulator MONACO, developed in our group, to discuss the evolution of the velocity distribution along the channel of a DGMOS. Their work puts into evidence that some electrons can be reflected towards the channel after having experienced Coulomb interactions in the highly doped drain region i.e. far from the low field region.

To analyze in more details the transport in nanometer-scaled DGMOS, we study in this paper the behavior of electrons injected from the source as a function of the number of scattering events undergone in the channel. We have introduced in the device Monte Carlo simulator a procedure which allows us to get accurate information on the number of scattering events experienced by each electron during its travel between the source-end and the drain-end of the channel. We can also extract at many positions in the channel the number of back-scattered electrons and the velocity distributions of electrons having undergone either 0, 1, 2, … or N scattering events. We thus have relevant information to investigate in detail in this paper the notion of ballisticity [6] and to discuss the actual influence of both ballistic and backscattering effects on the device characteristics and performance.

The paper is organized as follows. The simulated devices and the Monte Carlo model are presented in Section 2 and the results are described and discussed in section 3. First, we analyze the velocity spectra in the channel by highlighting the ballistic and quasi-ballistic phenomena, which completes the work initiated in Ref.[9]. Then the notion of intrinsic ballisticity is defined and its influence on drive current is evaluated by tuning the scatterings intensity in the channel. Finally, we investigate the evolution of backscattering coefficient along the channel and, by separating the channel in two parts with different scattering properties, we analyze the impact of the region where scattering events take place on the I-V characteristics.

## 2 Model and simulated devices

The simulated DGMOS devices, described in Fig. 1, have an effective channel length $L_{ch}$ equal to 10 nm, 15 nm, 25 nm or 50 nm and a gate lengths $L_G$ equal to 10, 25, 25 and 50 nm respectively. However, the study is focused on 15 nm-and 25 nm-effective channel lengths. The $SiO_2$ gate oxide $T_{ox}$ and Si body thicknesses $T_{Si}$ are equal to 1.2 nm and 10 nm, respectively. The doping density is $N_D = 5 \times 10^{19}$ cm$^{-3}$ in N$^+$ S-D regions and $N_A = 2 \times 10^{15}$ cm$^{-3}$ in the body (P type). The N$^+$ doping level is relatively low, and may induce a significant series resistance, but it is a realistic value considering the difficulty of electrically activating dopants in such thin body. The N$^+$/P junctions are assumed to be abrupt. The work function of the gate material is 4.46 eV to achieve the theoretical threshold voltage $V_T$ of 0.2 V. The power supply voltage $V_{DD}$ is fixed at 0.7 V to abide by the 2007 ITRS specification.

A classical particle Monte Carlo algorithm is self-consistently coupled with a 2D Poisson solver. The Poisson's equation is solved at each time step equal to 0.1 fs with standard boundary conditions [10]. The number of simulated particles is typically 50000. The scattering mechanisms included in the simulation are phonon scattering, impurity scattering and surface roughness scattering. The acoustic intra-valley phonon scattering is treated as an elastic process and the intervalley phonon transitions, consisting of three f-type and three g-type processes, are considered via either zeroth-order or first-order transition matrix in agreement with selection rules [11]. The phonon coupling constants given in Ref.[10] are used. The impurity scattering rate is derived from the screened Coulomb potential with the momentum-dependent screening length given in Ref.[12]. The surface roughness scattering is treated with an empirical combination of diffusive and specular reflections which correctly reproduces the experimental universal mobility curve [13],[14]. Unless otherwise indicated we have considered in this work a fraction of diffusive reflections $C_{diff}$ equal to 0.14.

The originality of the present work, in which quantum confinement effects are not included, lies in the spectroscopy of the number of scattering events undergone by electrons crossing the active part of the channel (or any predefined part of the device) and in the detailed study of the velocity spectra of different electron groups. This makes possible the careful analysis of scattering effects. Practically, a scattering counting region is predefined with an entrance surface and an exit surface. Typically, the entrance surface is defined either at the source/channel junction or at the position of the top of the gate-induced potential barrier, and the exit surface is placed at the channel/drain junction. Each electron entering the counting region by the entrance surface is flagged and while it remains inside this region the number of scattering events undergone is recorded. At the exit surface we can thus separate electrons into different groups, corresponding to ballistic electrons, once-scattered electrons, twice-scattered electrons, … N-times scattered electrons, respectively. Of course these groups can be enumerated. Additionally, for each of these groups the energy and velocity spectra are recorded. Such spectra may be also obtained at intermediate surfaces defined all along the counting region, i.e. the channel. It should be noted that these spectra only include electrons coming from the entrance surface and do not consider the electrons entering the counting region by the exit surface. The information about back-scattered electrons is obtained separately: if an electron is back-scattered after having crossed the counting region and re-enter this region by the exit surface, its scattering-story is still recorded and the velocity spectra of such electrons are recorded too. Thus, by possibly changing the place of the exit surface, we can have very detailed information on all carriers participating in the source-drain current. It is very useful to quantify and analyze the effects of ballistic, nearly-ballistic and back-scattered electrons.

## 3  Results

The variations of drain current $I_D$ as a function of source-drain voltage $V_{DS}$ obtained in the 15 nm-, 25 nm- and 50 nm-long transistors for $V_{GS} = V_{DD}$ are shown in Fig. 2. For the 15 nm-long channel, the $I_{on}$ current at $V_{DS} = V_{GS} = V_{DD} = 0.7$ V is 2140 µA/µm. Such a high value is related to the aggressive scaling of the gate length, to the double-gate architecture and to the relatively thick body film ($T_{si} = 10$ nm). The drain conductance, and more generally the short channel effects (SCE), are however rather strong in the 15 nm-long channel, as a consequence of the non optimized body thickness. The 25 nm-long channel, better designed, is less sensitive to SCE while exhibiting $I_{on}$-value (1600 µA/µm) still higher than the ITRS Roadmap specification that is 900 µA/µm. We now examine the transport in these devices biased in the "on-state", i.e. with $V_{GS} = V_{DS} = V_{DD}$.

## 3.1 Velocity spectra analysis

The conduction band profile and the corresponding velocity spectra evolution of electrons along the 15 nm-long channel have been plotted respectively in Fig. 3 and Fig. 4. The Fig. 4 represents the velocity spectra evolution calculated in the first sheets of cells under the front oxide, i.e. between y = 0 and y = 1 nm (see Fig. 1). The transport is stationary in the highly doped source well, as illustrated by the quasi-Gaussian shape of the velocity distribution in the middle of this region. Other spectra are taken in the channel and, in contrast to the results shown in Ref.[9], they only concern electrons injected by the source into the channel, excluding those which have been injected or re-injected from the drain. These spectra are very different from stationary ones. In the vicinity of the top of the barrier (x = 2 nm), the spectrum is quite similar to a hemi–Gaussian distribution. However electrons with a negative velocity have not completely disappeared, which is the signature of backscattered electrons. Thus, the regime of pure thermionic injection in the channel is not completely reached in this structure. At a position x = 4 nm after the source/channel junction, two separate distributions appear in the velocity spectrum. Two peaks can be distinguished, each of them corresponding to electrons with either a transverse or a longitudinal conductive effective mass along the source-to-drain direction. For greater distance x, we observe the propagation of these two distributions along the channel, each one at its own velocity. The electron density decreases as the average velocity increases in accordance with the current conservation and the distributions become narrower and narrower.

At the drain-end of the 15 nm-long device, the velocity spectrum of electrons injected from the source calculated at the $Si/SiO_2$ interface looks like two very well-defined peaks as shown in Fig. 5. According to energy conservation between the barrier region and the drain-end of the channel, it is easily demonstrated that the peak velocity values in Fig. 5 correspond to ballistic electrons flowing from source to drain with a transverse ($m_t = 0.19\ m_0$ where $m_0$ is the free electron mass) or a longitudinal ($m_l = 0.916\ m_0$) effective mass.

To complete the description of electron populations present in the channel, the spectrum of electrons injected from the drain at x = 15 nm is also plotted in dotted line in Fig. 5. As confirmed by a velocity spectroscopy distinguishing electrons with transverse mass from those with longitudinal mass (not shown), this 'drain injected' distribution is made up of the sum of two nearly Gaussian distributions. The widest corresponds to electrons with a transverse mass and the thinnest to electrons with a longitudinal mass. Even if they are the most numerous at the drain-end, their net contribution to the current at the drain-end is only 0.9%.

Besides, the number of "drain-injected" electrons rapidly decreases near the source. Indeed, at x = 15, 12, 4 and 1 nm, the part of "drain injected" electrons represents respectively 72.3%, 42.3%, 5.6% and 1.7 % of the electrons. At the source-end (x = 1 nm) 6.3% of electrons which have a negative velocity have been injected from the highly-doped drain region.

Those first analyses of velocity evolution along the channel indicate clearly that the transport is completely out of equilibrium. Furthermore, the velocity spectrum of "source-injected" electrons at the drain-end (Fig. 5) suggests that ballistic electrons have an important role in the transport for nanometer-scaled DGMOS devices. In this connection, we present a new kind of analysis which allows dissecting the velocity spectra presented in Fig. 5. In this purpose, we have defined a counting region by an entrance surface and an exit surface located at the source/channel and channel/drain junctions, respectively. This study does not consider "drain-injected" electrons.

The Fig. 6 represents the velocity distributions calculated at the drain-end of the 15 nm-long channel device for electrons flowing from the source by undergoing 2, 1 or 0 scattering events in the counting region. Contrary to the spectra presented above (i.e. in Fig. 4 and Fig. 5) and calculated at the Si/SiO$_2$ interface, these new distributions are calculated on the full body thickness, i.e. between y = 0 and y = T$_{si}$ (cf. Fig. 1). A 2D effect is clearly observed in Fig. 6: the ballistic peaks are wider than those presented in Fig. 5 as they result from the sum of peaks with different maxima. Indeed, as shown in Fig. 3, the potential drop varies as a function of the channel depth, i.e. along the y-axis (cf. Fig. 1), which makes the ballistic peak position dependent on y. For instance, according to the potential drop plotted in Fig. 3, the transverse mass peak velocity is equal to 6.5 × 10$^7$ cm/s at y = 0 nm and to 7.0 × 10$^7$ cm/s at y = T$_{si}$/2. We also verify that the distribution tails observed in Fig. 5 are caused by electrons which have undergone interactions during their channel crossing. Moreover, the shapes of the quasi-ballistic velocity spectra, i.e. the spectra obtained for electrons which have undergone 1 or 2 interactions, is similar to that of a ballistic spectrum. This suggests that devices can drive on-current I$_{on}$ very close to the limit value obtained for a ballistic channel, even if all electrons are not purely ballistic at the drain-end.

The Fig. 7 represents the same kind of velocity distributions for "source injected" electrons but calculated near the top of the injecting barrier, i.e. at x = 3 nm from the source-end. This figure allows us to verify that only ballistic electrons are purely thermionically injected since their velocity spectrum looks like a strict hemi-Gaussian distribution. Additionally, the contribution of electrons which have undergone more than two scatterings in the channel is negligible (not shown). At V$_{DS}$ = 0.7 V, the backscattered population is essentially formed by once– and twice–scattered electrons.

### 3.2 Influence of the ballisticity on drive current

To get a more quantitative insight into ballistic and/or quasi-ballistic transport, we have calculated the number of scattering events $N_{scatt}$ experienced by each carrier crossing the counting region, i.e. the channel. The resulting electron distribution is plotted in Fig. 8 as a function of $N_{scatt}$ for DGMOS of different effective channel lengths: 15 nm, 25 nm and 50 nm. For comparison, we have also indicated, in dotted line the similar distribution obtained at the drain-end for a conventional 50 nm bulk MOSFET with the following characteristics: a single gate, a channel length $L_{ch} = 50$ nm, an oxide thickness $T_{ox} = 1.2$ nm, a junction depth $X_j = 20$ nm and a highly doped channel $N_A = 10^{18}$ at/cm$^3$. The latter distribution is a bell-curve with a maximum centered on $N_{scatt} = 5$, which corresponds to the ratio $L_{ch}/\lambda_{eff}$ where $L_{ch}$ is the effective channel length and $\lambda_{eff}$ is defined as an effective mean free path. On the contrary, in all lightly-doped DGMOS the group of ballistic electrons is the most populated, but it forms the majority only in the 15 nm-long channel. This indicates that the effective mean free path $\lambda_{eff}$ is smaller than the channel length in all simulated DGMOS. However, the fraction of electrons decreases as $N_{scatt}$ (the number of scatterings) increases and the distribution spreads out when $L_{ch}$ increases. For the 15 nm-long channel, the curve is a pure exponential function. In both 50 nm-long channel devices (DGMOS and bulk) the interaction spectrum tends to a more stationary-like one.

Now, we define the intrinsic ballisticity $B_{int}$ as the percentage of electrons which are ballistic at the drain-end ($N_{scatt} = 0$ in Fig. 8) [15]. The line in Fig. 9 is an interpolated curve obtained by linking the intrinsic ballisticity $B_{int}$ for the 3 different DGMOS with channel effective length $L_{ch}$ equal to 15 nm, 25 nm and 50 nm and by assuming $B_{int} = 100\%$ for $L_{ch} = 0$. The smooth curve obtained from these three DGMOS seems particularly relevant because for $L_{ch} = 10$ nm it gives the same result as the complete Monte Carlo simulation of a 10 nm-long device (closed circle). From such a semi classical Monte Carlo approach, one can thus estimate that ballistic electrons should be largely predominant (i.e. with an intrinsic ballisticity $B_{int}$ greater than 90%) only for channel lengths smaller than about 3 nm. However, quantum transport effects have to be considered for investigating $L_{ch}$-values smaller than 10 nm [7], which may modify this prediction.

To determine how the intrinsic ballisticity $B_{int}$ at the drain-end is related to the on-current $I_{on}$, we have artificially modified it from 0 as in full stationary transport to 100% as in pure ballistic transport. The quantity $I_{on\_bal}$ stands for the on-current obtained for a ballistic channel, that is without any phonon or roughness effects. Then, we study the effective ballisticity $B_{eff}$, defined as $B_{eff} = I_{on}/I_{on\_bal}$ [6], as a function of the intrinsic ballisticity $B_{int}$ at the drain-end.

To this end, we have varied the oxide roughness coefficient $C_{diff}$ from 0 to 1 and introduced a phonon scattering coefficient $K_{ph}$ in the phonon scattering rates: all standard values are multiplied by the coefficient $K_{ph}$ varying from 0 for a ballistic channel, to 20 for a very resistive channel. The evolution of $B_{eff}$ as a function of $B_{int}$ for a given $C_{diff}$ and for different phonon scattering coefficients, is plotted in dotted line in Fig. 10. The results obtained for $C_{diff} = 1$ (respectively $C_{diff} = 0.14$) and for $K_{ph} = 0, 1, 2, 5, 10$ and 20 (respectively $K_{ph} = 0, 0.5, 1, 2$), are indicated with open squares (respectively closed triangles). The results obtained for various roughness coefficients $C_{diff}$ and for a given phonon coefficient $K_{ph}$ are shown in solid lines: on the one hand for $K_{ph} = 1$ and $C_{diff} = 0, 0.07, 0.14, 0.21, 0.4, 0.7$ and 1 (open circles), and, on the other hand, for $K_{ph} = 0$ with $C_{diff} = 0, 0.14, 0.5$, and 1 (closed squares).

These results show an overall view of the effect of each kind of interaction. For an intrinsic ballisticity $B_{int}$ greater than 20% the effect of each type of interaction (phonon or roughness scattering) yields a linear behavior $B_{eff}(B_{int})$ but with a slope depending on the type of interaction. Below this limit, when the transport is more stationary, the $B_{eff}(B_{int})$ relation is no more simply linear. Thus there is neither an equality nor a unique linear relation between the effective ballisticity $B_{eff}$ and the intrinsic ballisticity $B_{int}$. The effective ballisticity $B_{eff}$ (in terms of current) alone does not provide enough information to quantify accurately the intrinsic ballisticity $B_{int}$. However, there is no denying that these two quantities are strongly correlated. Besides, we notice that the intrinsic ballisticity $B_{int}$ is always overestimated by the effective ballisticity $B_{eff}$: for instance $B_{int} = 52\%$ and $B_{eff} = 84\%$ for the standard DGMOS ($K_{ph} = 1$ and $C_{diff} = 0.14$).

## 3.3 Back-scattering localization and its effect on the current

To investigate further the effects of scattering, a flux approach may be used [6]. We propose to identify the part of the channel giving the highest contribution to back-scattering effects which are known to degrade the drive current. So, a control volume is defined, as shown in Fig. 11, by an entrance surface and an exit surface that can be moved along the channel. By calculating, the different fluxes: $\Phi_I^+$, $\Phi_O^+$, $\Phi_{I1}^-$ and $\Phi_{I2}^-$, the relevant back-scattering coefficients may be extracted. Fluxes $\Phi_I^+$ and $\Phi_O^+$, oriented as indicated in Fig. 11, are the ingoing flux at the entrance surface and the outgoing flux at the exit surface, respectively. The flux $\Phi_{I1}^-$ represents the flux of electrons which have entered the control volume by the entrance surface and which have crossed back the entrance surface without having crossed the exit surface. The flux $\Phi_{I2}^-$ represents the flux of electrons which have entered the control volume by the entrance surface and which have crossed back the entrance surface after having crossed the exit surface.

The back-scattering coefficient R(x) of the control volume between the entrance surface located at the top of the barrier and an exit surface located at the distance x from the source end is equal to $\Phi_{I1}^-(x)/\Phi_I^+$. The evolution of R(x) obtained by moving the exit surface along the channel from the top of the barrier to the drain–end is plotted in Fig. 12 for 15 nm- and 25 nm-long devices. The maximum back-scattering coefficient $R_{max}$ for the whole structure including the drain contact can be calculated as $R_{max} = \max[R(x)] = [\Phi_{I1}^-(x) + \Phi_{I2}^-(x)]/\Phi_I^+$.

First, we remark that at the drain-end of the 25 nm–long channel the backscattering coefficient $R(L_{ch})$ is slightly lower than $R_{max}$: $R_{max} = 1.035 \times R(L_{ch})$. Moreover, the difference between $R_{max}$ and $R(L_{ch})$ strongly increases when $L_{ch}$ decreases to 15 nm: $R_{max} = 1.195 \times R(L_{ch})$. This difference, due to electrons back-scattered in the drain, may be more accurately estimated by taking into account other scattering mechanisms, as for instance, the short range electron-electron interactions.

A monotonous variation of R(x) along the channel is observed. The increase of R(x) is much more important in the first half part of the channel as the back-scattering events take place mainly there. However this is less true when the effective channel length $L_{ch}$ decreases: $R(x = L_{ch}/2) = 0.91 \times R(L_{ch})$ for $L_{ch} = 25$ nm, while $R(x = L_{ch}/2) = 0.81 \times R(L_{ch})$ for $L_{ch} = 15$ nm.

In the first part of the channel the increase of R(x) is rather uniform, just a bit faster in the vicinity of the top of the barrier. Thus, it is difficult to accurately define in the first channel half a region which has a predominant impact in terms of back-scattering.

It has been suggested in Ref. [6] that the backscattering coefficient R can be estimated at the position x in the channel where the potential drops by kT/q. Our results show that this definition leads to a significant underestimation of R. For instance, consider the position $x_0$ where the potential drops by at least 50 meV on the full channel thickness. We have then $x_0 = 9.5$ nm for $L_{ch} = 25$ nm and $x_0 = 6.3$ nm for $L_{ch} = 15$ nm. At this position $x_0$, the backscattering coefficient is only $R(x_0) = 0.86 \times R(L_{ch}) = 0.78 \times R_{max}$ for $L_{ch} = 25$ nm. This underestimation is more pronounced for $L_{ch} = 15$ nm: $R(x_0) = 0.73 \times R(L_{ch}) = 0.61 \times R_{max}$.

This trend confirms that reducing $L_{ch}$ gives more importance to the second half of the channel in terms of contribution to backscattering. Indeed, the number of scattering in ultra-short channel is not important enough (cf. Fig. 8) to prevent electrons scattered in the second half of the channel from being backscattered to the source–end [8].

To analyze the influence on drain current of scattering in the different parts of the channel, new simulations have been performed by changing the scattering properties along the channel. Three new 25 nm-long DGMOS called 'bal-bal', 'bal-sta' and 'sta-bal' have been simulated. In these devices, the channel is divided in 2 equal parts. The term 'sta' stands for the channel–half with standard scattering properties of Si doped to $2 \times 10^{15}$ at/cm$^3$: standard phonon scattering coefficient $K_{ph} = 1$ (see Sec. 3.2) and roughness coefficient $C_{diff} = 0.14$. The term 'bal' stands for the ballistic channel–half (without any scattering). So, for instance, 'sta-bal' is a DGMOS with a standard first channel–half and a ballistic second channel–half.

The inset of Fig. 13 shows on both linear and logarithmic scales the evolution of drain current $I_D$ as a function of gate voltage $V_{GS}$ at $V_{DS} = V_{DD}$. First we notice that all devices have the same threshold voltage $V_T \approx 0.3$ V. Moreover, the subthreshold behavior is not degraded by SCE, the subthreshold slope being equal to 70 mV/dec. The characteristics only differ in the transconductance $g_m = \left. \frac{\partial I_D}{\partial V_{GS}} \right|_{high\ V_{GS}}$ above the threshold voltage $V_T$ and, as a consequence, in the on-current $I_{on}$.

The variations of drain current $I_D$ as a function of source-drain voltage $V_{DS}$ obtained in these DGMOS at $V_{GS} = 0.7$ V are shown in solid lines in Fig. 13. In linear regime, the resistance $R_{on} = \left. \frac{\partial V_{DS}}{\partial I_D} \right|_{low\ V_{DS}}$ extracted from Fig. 13 simply follows the Ohm's law: $R_{on}$ is determined by the average channel conductivity and has the same value for 'bal-sat' and 'sta-bal'. It is greater for the standard channel and lower for the ballistic channel. In the saturation regime, with nearly the same saturation drain voltage $V_{DSsat}$ for all devices, the following values of output conductance $G_D = \left. \frac{\partial I_D}{\partial V_{DS}} \right|_{V_{GS}=0.7V}$ are extracted at $V_{GS} = 0.7$ V: $G_{D\_BB} = 360$ μS/μm for 'bal-bal', $G_{D\_BS} = 340$ μS/μm for 'bal-sta', $G_{D\_SB} = 270$ μS/μm for 'sta-bal', $G_{D\_SS} = 190$ μS/μm for 'standard'. We observe that the more ballistic the channel is, in particular in its first half, the greater the conductance is. Thus, we deduce that the lack of scattering degrades the saturation behavior. It should be mentioned that similar results have been obtained for bulk MOSFET of higher channel length, i.e. 50 nm and 180 nm (not shown). Low $R_{on}$, i.e. high channel conductivity, is beneficial for the on-current $I_{on}$, and a weak conductivity, in particular the first channel–half, is suitable to get a low output conductance $G_D$.

Besides, if each channel part had the same influence on drive current, 'sta-bal' and 'bal-sta' on–current would be equal to the average on-current between 'standard' and 'bal-bal', i.e. equal to 1980 μA/μm. As the on-current $I_{on}$ is equal to 2000 μA/μm for 'bal-sta' and 1850 μA/μm for 'sta-bal', we deduce that the ballistic channel part location, i.e. the scattering localization, has an influence on the on-current.

This result is not fully inconsistent with the conventional belief that the first channel-half has a greater importance than the second channel-half in terms of backscattering coefficient. However, in contrast with this belief, our study reveals that the second part of the channel has a significant impact on device operation and performance. Indeed $I_{on}$ is 15% greater for 'sta-bal' than for 'standard' and it is 14% greater for 'bal-bal' than 'bal-sta'. It should be mentioned that the intrinsic ballisticity is higher for 'sta-bal' (61%) than for 'bal-sta' (53%), the resulting $I_{on}$ is higher for the latter device. Moreover the relative difference between 'sta-bal' and 'bal-sta' is low: $I_{on}$(bal-sta) - $I_{on}$(sta-bal) = 4.4% × ($I_{on}$(standard) - $I_{on}$(bal-bal)). At last, the evolution of the backscattering coefficient along the channel (Fig14) suggests that this impact of the second channel–half should be still stronger for smaller channel length.

## 4    Conclusion

We have investigated in detail the velocity spectra of the electrons present in the channel as a function of their origin: "source-" and "drain-injected" and of the number of the experienced scatterings. We show that the transport in the 15 nm-long channel is neither purely ballistic nor purely thermionic.

Nevertheless, ballistic electrons are of great importance in nanometer-scaled double-gate MOSFET. In nano-DGMOS, the ballistic limit (intrinsic ballisticity $B_{int}$ = 100% i.e. 100% of electrons injected from the source are ballistic at the drain-end) is far to be reached for channel lengths larger than 10 nm: the intrinsic ballisticity is about 50% in 15 nm-long channel. However, the drive current $I_{on}$ is closer to the value $I_{on\_bal}$ obtained with a pure ballistic channel: the ratio $I_{on}/I_{on\_bal}$, i.e. the effective ballisticity $B_{eff}$, is more than 80%. Then the number of ballistic electrons, even in terms of flux as we defined the intrinsic ballisticity $B_{int}$, is not the only relevant characteristic of ballistic transport. There even if the "ballistic limit" is still "a mere pipe dream" considering the numerous types of significant interactions in nano-scaled structures [16], the on-current in next transistor generations should be very close to this limit.

Even if our results are not fully inconsistent with the conventional belief that most back-scattering take place in the first half of the channel, they show that the role of the second half of the channel cannot be considered as negligible for a channel length lower than 25 nm. Furthermore the contribution of the second half to back-scatterings tends to be more and more important as the channel length is reduced. As a consequence, it becomes more and more difficult to extract a particularly significant region of the channel which would determine the value of the drive current $I_{on}$.

This work is supported by the French RMNT project CMOS-D-ALI and we thank Emile Grémion for his contribution.

**Figure captions**

Fig. 1: Schematic of DGMOS structures.

Fig. 2: Drain current $I_D$ versus drain voltage $V_{DS}$ at $V_{GS} = 0.7$ V in DGMOS of different channel lengths $L_{ch}$. Crosses: $L_{ch} = 50$ nm, closed circles: $L_{ch} = 25$ nm, and open squares: $L_{ch} = 15$ nm.

Fig. 3: Conduction band versus distance x along the S-D direction in the on-state ($V_{GS} = V_{DS} = V_{DD}$) for the 15 nm-long transistor: $L_{ch} = 15$ nm and $L_G = 25$ nm for different distances y from the gate: y = 0 nm for the Si/SiO$_2$ interface and y = $T_{Si}$ / 2 for the body center. The dotted lines indicate the S-D gate overlap, the x-axis origin corresponds to the position of the source/channel junction.

Fig. 4: Velocity $v_x$ distributions calculated in the N$^+$ source region and at different positions into the 15 nm-long channel. For x>0, only electrons injected from the source are considered.

Fig. 5: Velocity $v_x$ distributions of electrons flowing from the source (solid lines) and electrons flowing from the drain (dotted lined) calculated at the drain-end (x = 15 nm) of the 15 nm-long channel.

Fig. 6: Velocity $v_x$ distributions calculated at the drain-end (x = 15 nm) of the 15 nm-long channel for electrons flowing from the source and undergoing either 2, 1 or 0 scattering events.

Fig. 7: Velocity $v_x$ distributions calculated at x = 3 nm from the source-end of the 15 nm-long channel for electrons flowing from the source, and undergoing either 2, 1 or 0 scattering events.

Fig. 8: Fraction of electrons flowing from S to D versus the number of scattering events $N_{scatt}$ undergone during the channel crossing. Conventional bulk MOSFET (Single gate, $L_{ch} = 50$ nm) distribution: dotted line. DGMOS distributions: solid lines. Crosses: $L_{ch} = 50$ nm, closed circles: $L_{ch} = 25$ nm, and open squares: $L_{ch} = 15$ nm

Fig. 9: Intrinsic Ballisticity $B_{int}$ at the drain-end versus channel effective length $L_{ch}$.

Fig. 10: Effective ballisticity $B_{eff} = I_{on} / I_{on\_bal}$ versus intrinsic ballisticity $B_{int}$ in the 15 nm-long device. Results are obtained by varying the intensity of the oxide roughness $C_{diff}$ or the phonon scattering coefficient $K_{ph}$. For $C_{diff} = 1$ (respectively $C_{diff} = 0.14$) and $K_{ph} = 0, 1, 2, 5, 10$ and 20 (respectively $K_{ph} = 0, 0.5, 1, 2$): open squares (respectively closed triangles). For $K_{ph} = 1$ (respectively $K_{ph} = 0$) and $C_{diff}$: 0, 0.07, 0.14, 0.21, 0.4, 0.7 and 1 (respectively $C_{diff} = 0, 0.14, 0.5$, and 1): open circles(respectively closed squares).

Fig. 11: Fluxes schematic and conduction band versus distance x along the S-D direction. Vertical dotted lines indicate source/channel and channel/drain junctions.

Fig. 12: Back-scattering coefficient R(x) versus S-D distance x in the channel. Closed circles: $L_{ch}$ = 25 nm and open squares: $L_{ch}$ = 15 nm.

Fig. 13: $I_D$ versus $V_{GS}$ at $V_{DS}$ = 0.7 V for different devices: 'standard' (closed circles), 'sta-bal' (open diamonds), 'bal-sta' (open circles), and 'bal-bal' (closed squares). Inset: $I_D$ versus $V_{GS}$ at $V_{DS}$ = 0.7 V

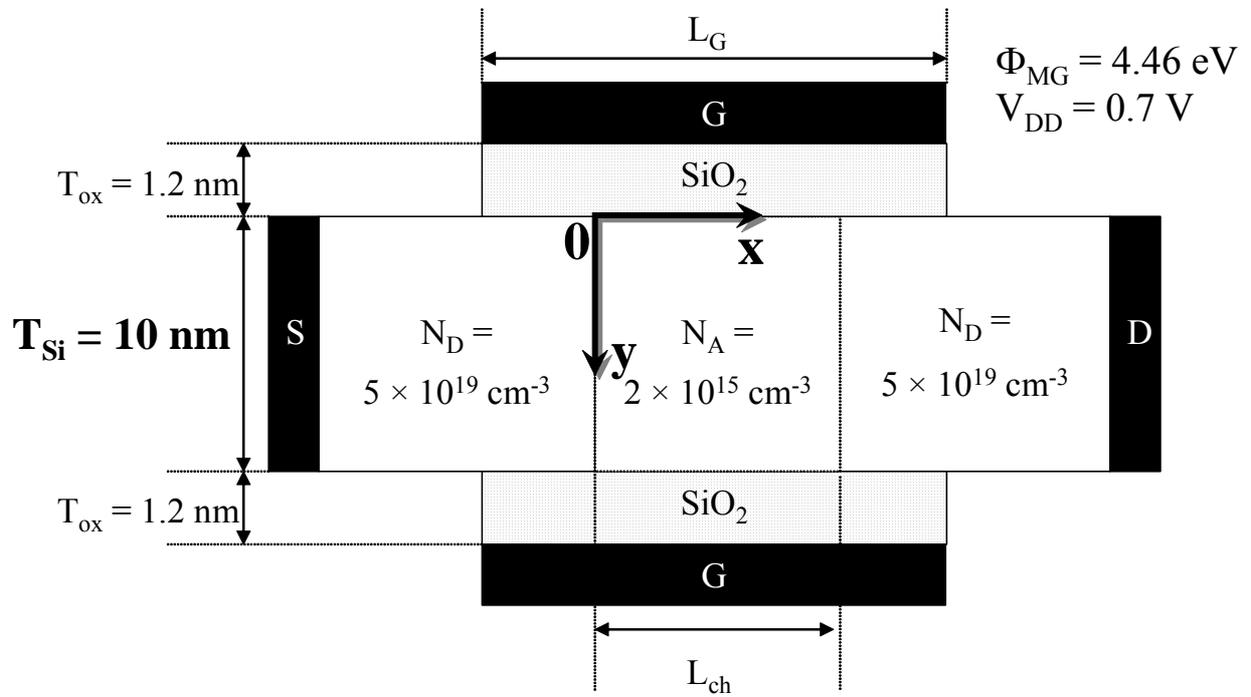

**Fig. 1**

Saint Martin *et al.*

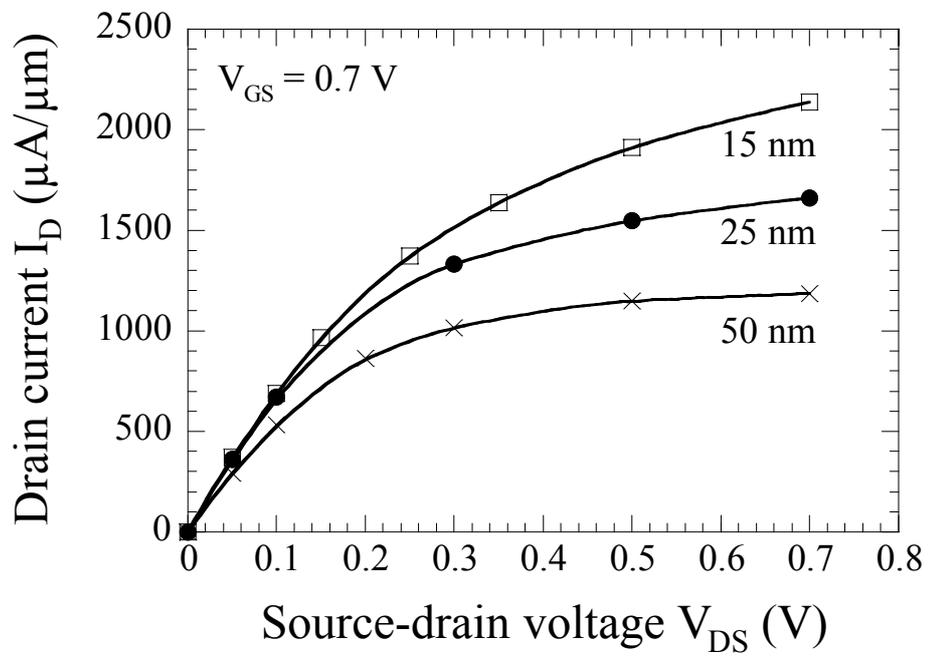

**Fig. 2**

Saint Martin *et al.*

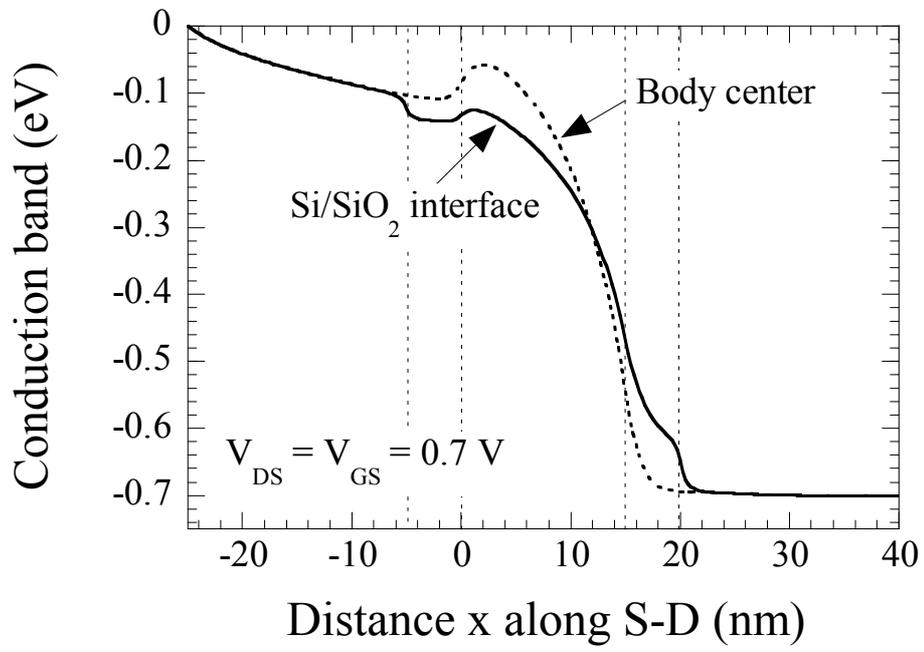

**Fig. 3**

Saint Martin *et al.*

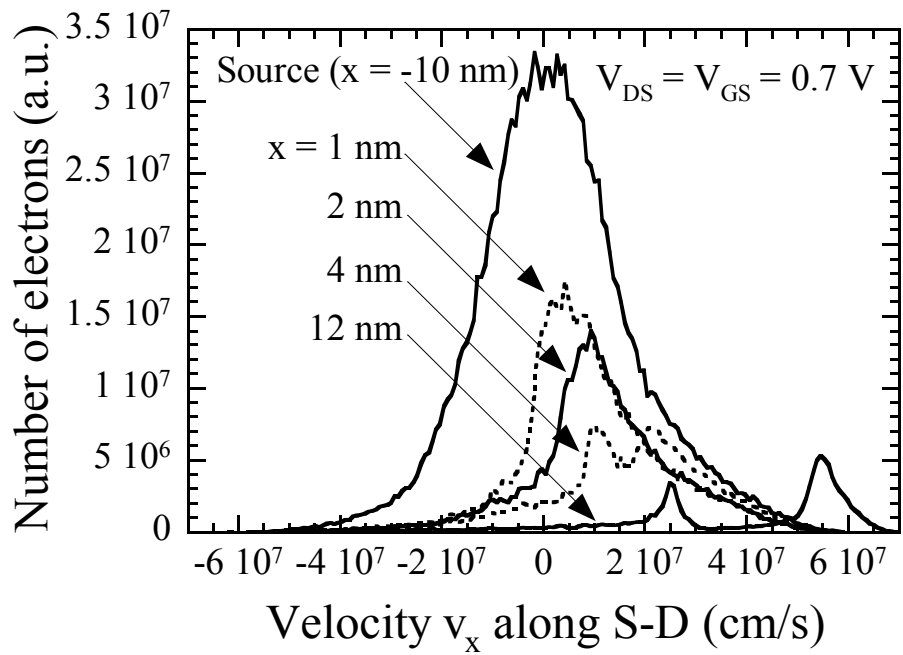

**Fig. 4**

Saint Martin *et al.*

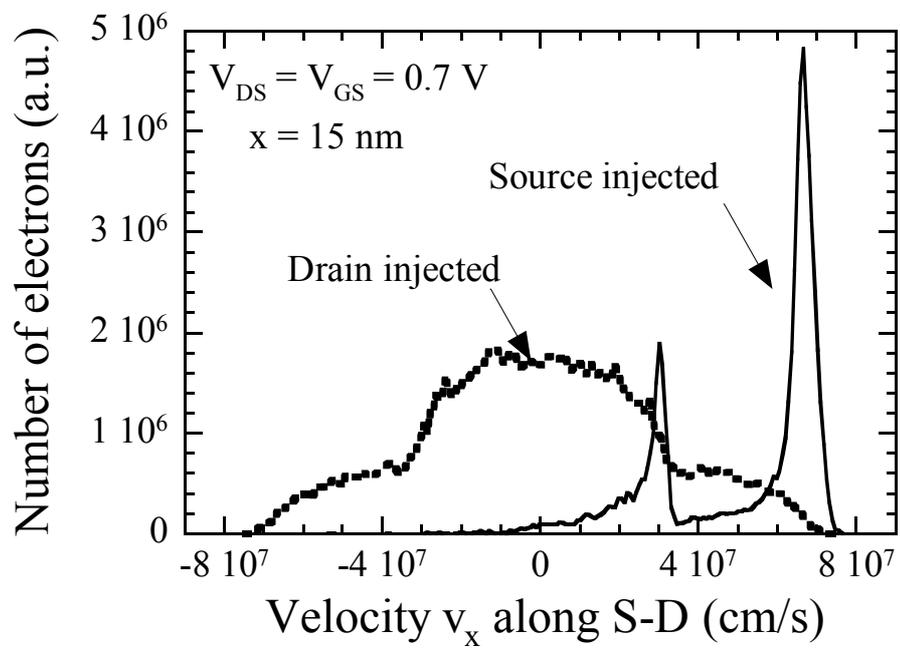

**Fig. 5**

Saint Martin *et al.*

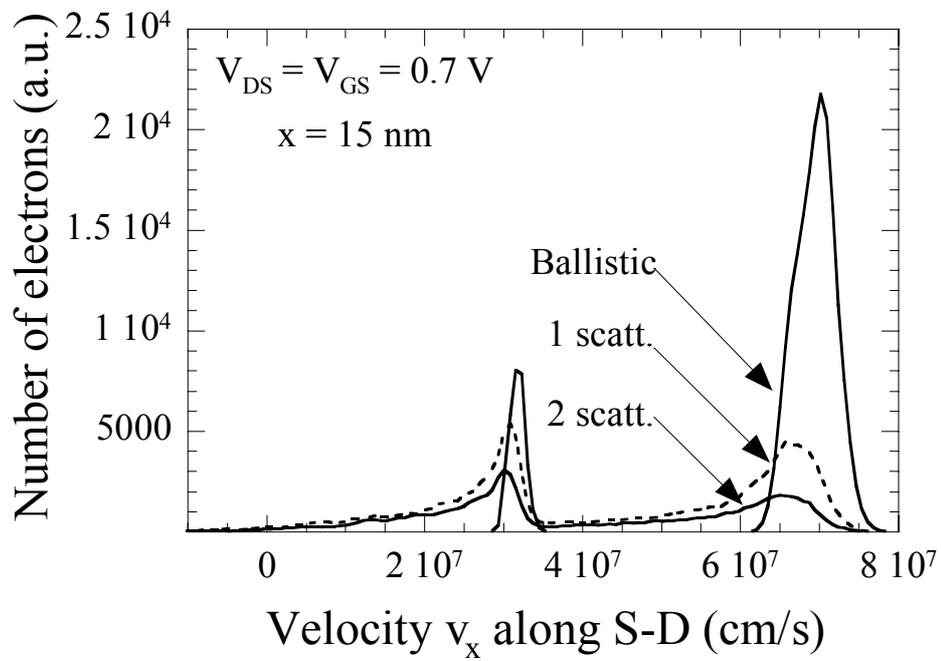

**Fig. 6**

Saint Martin *et al.*

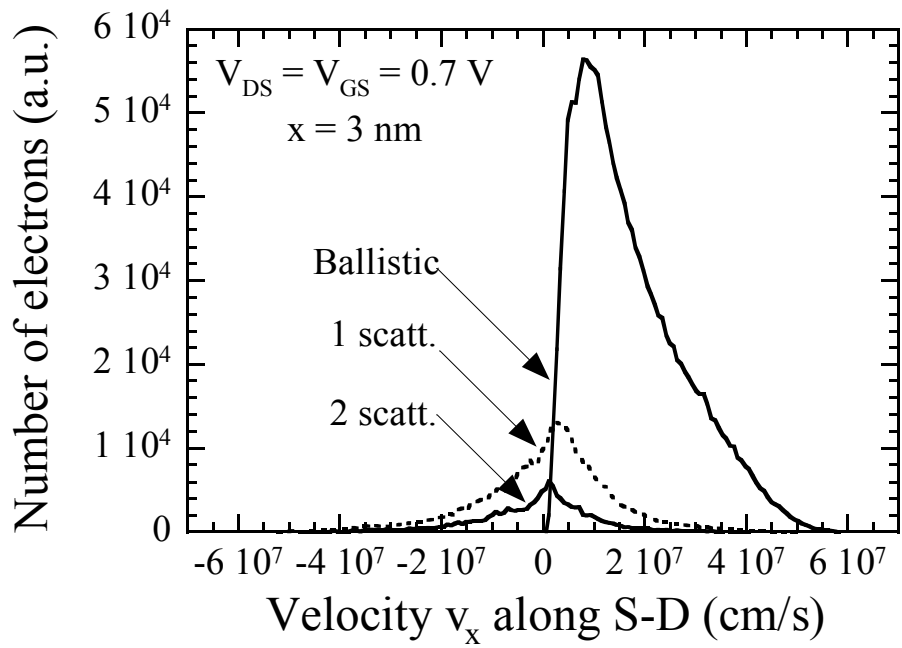

**Fig. 7**

Saint Martin *et al.*

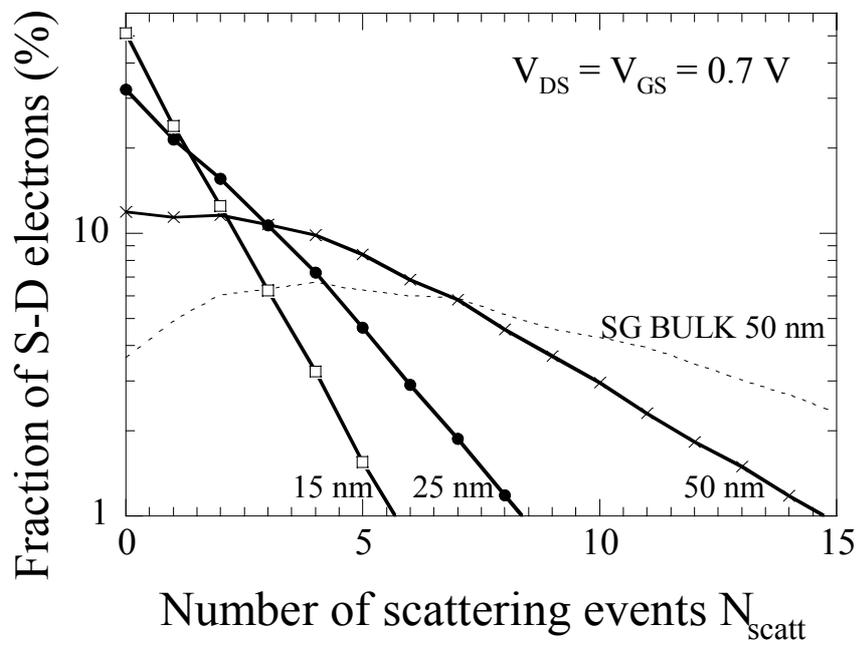

**Fig. 8**

Saint Martin *et al.*

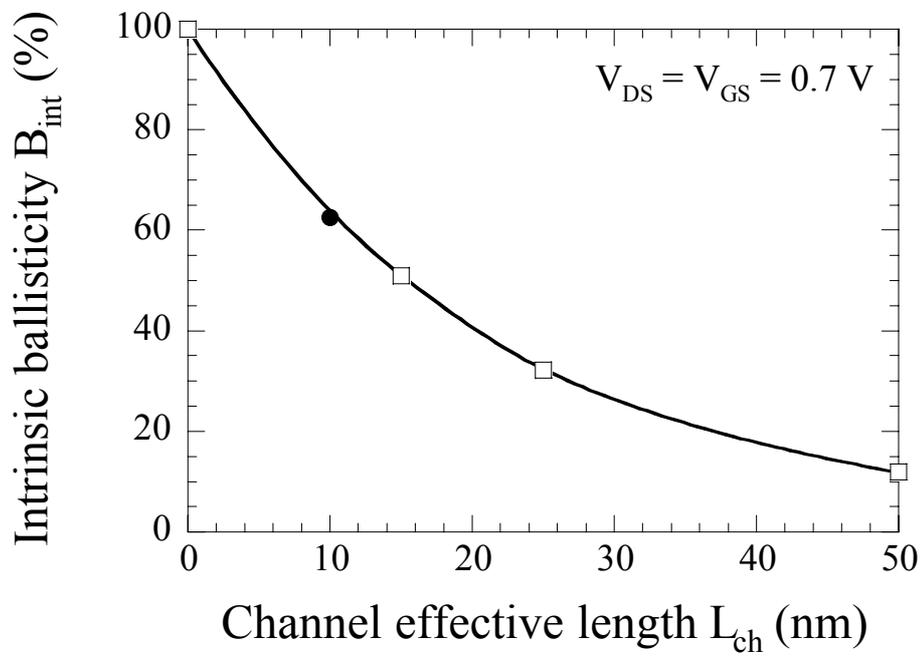

**Fig. 9**

Saint Martin *et al.*

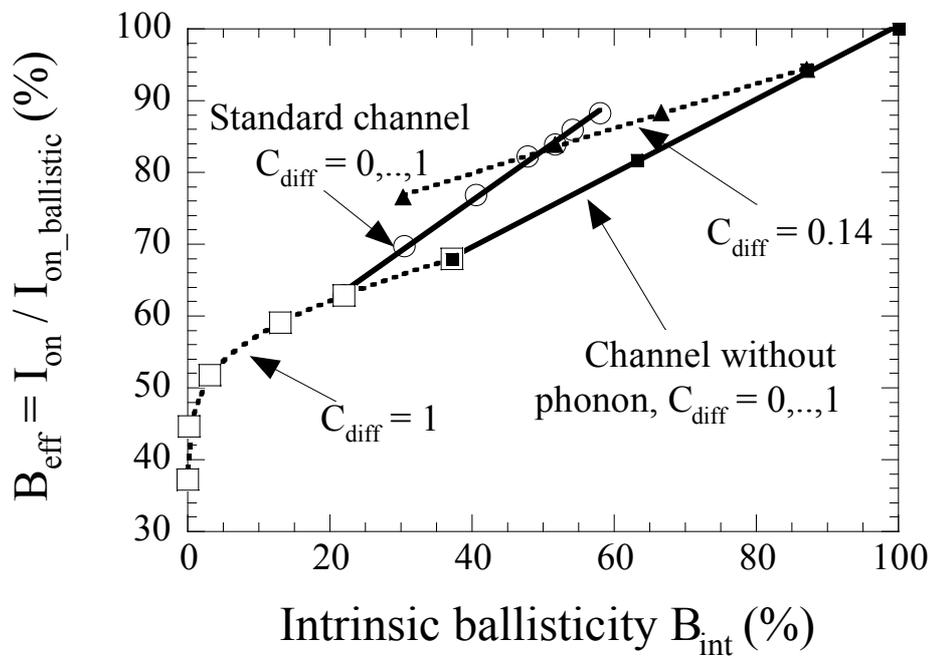

**Fig. 10**

Saint Martin *et al.*

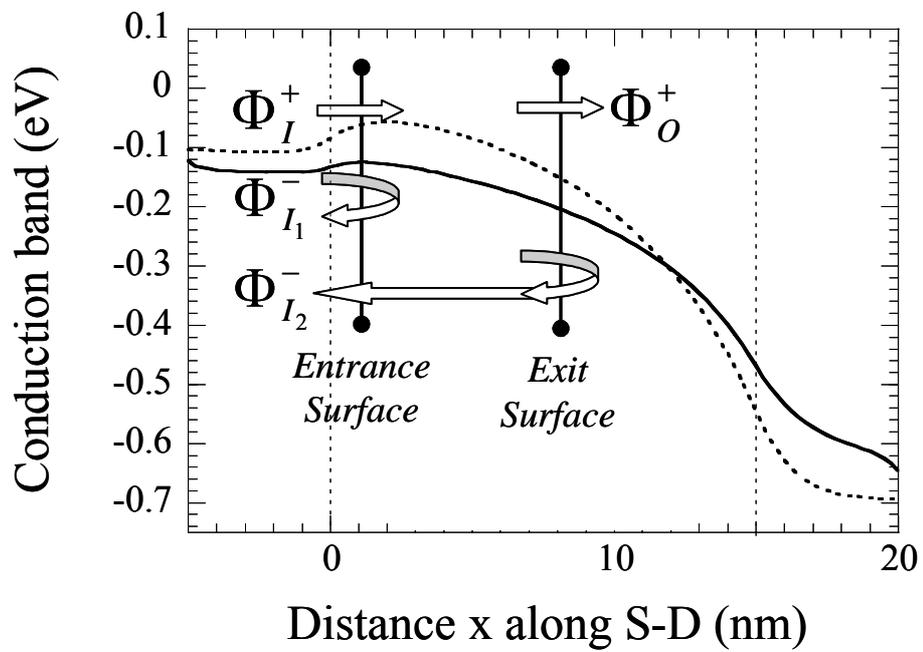

**Fig. 11**

Saint Martin *et al.*

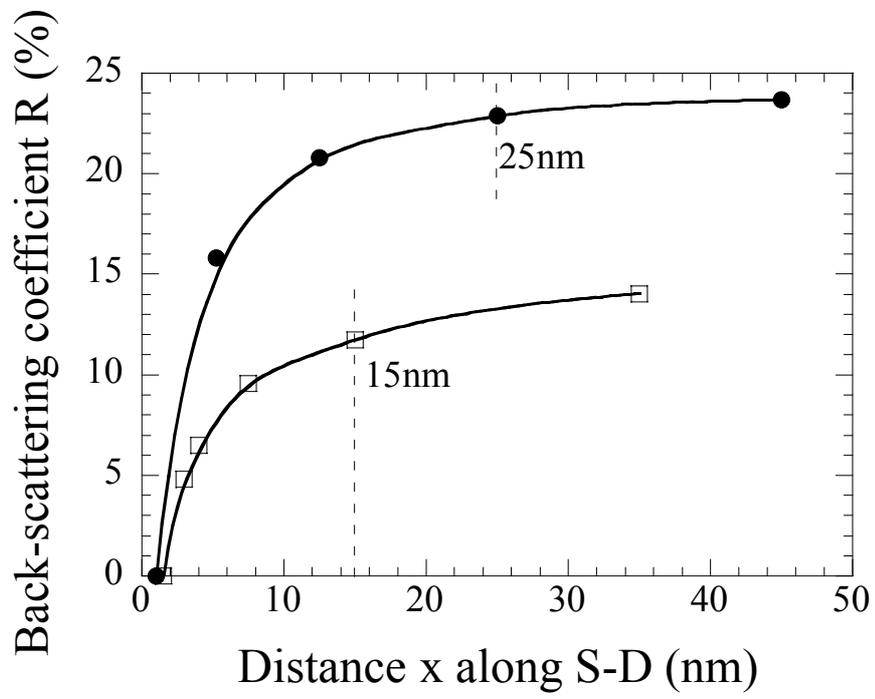

**Fig. 12**

Saint Martin *et al.*

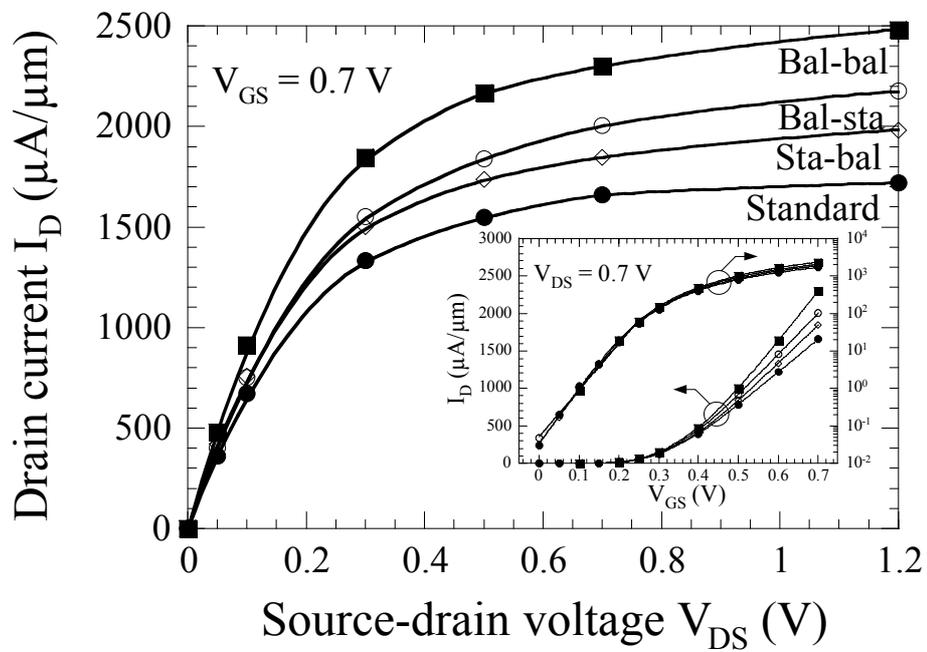

**Fig. 13**

Saint Martin *et al.*